\documentclass[smallextended]{svjour3}
\RequirePackage{fix-cm}
\usepackage[utf8]{inputenc}
\usepackage{epsfig}
\usepackage{amsmath}
\usepackage{graphicx}
\usepackage[draft]{hyperref}  
\usepackage{url}
\usepackage{wrapfig}

\newcommand{\mean}[1] {{\mathrm E}\left[#1\right]}
\newcommand{\cov}[2] {{\mathrm D}\left[#1, #2\right]}
\newcommand{\var}[1] {{\mathrm D}\left[#1\right]}
\newcommand{\arima}[3] {{${\mathrm {ARIMA}}\left(#1,#2,#3\right)$}}

\usepackage{hyperref}

\begin{document}
\title{Forecasting seeing and parameters of long-exposure images by means of ARIMA}
\author{Matwey V. Kornilov}
\institute{M. Kornilov \at
	Sternberg Astronomical Institute, Lomonosov Moscow State University, \\
	Universitetsky~pr.,~13, Moscow~119991, Russia \\
	\email{matwey@sai.msu.ru}
}

\maketitle

\begin{abstract}
Atmospheric turbulence is the one of the major limiting factors for ground-based astronomical observations.
In this paper, the problem of short-term forecasting seeing is discussed.
The real data that were obtained by atmospheric optical turbulence (OT) measurements above Mount Shatdzhatmaz in 2007–2013 have been analysed.
Linear auto-regressive integrated moving average (ARIMA) models are used for the forecasting.
A new procedure for forecasting the image characteristics of direct astronomical observations (central image intensity, full width at half maximum, radius encircling $80\%$ of the energy) has been proposed.
Probability density functions of the forecast of these quantities are 1.5–2 times thinner than the respective unconditional probability density functions.
Overall, this study found that the described technique could adequately describe temporal stochastic variations of the OT power. 

\end{abstract}

\section{Introduction}
\label{sec:intro}

Seeing is one of the most important parameters for determining the performance of ground-based astronomical observations.
Traditional definitions of telescope effectiveness for photometrical measurements usually involve the exposure time which is required to obtain given result.
In general, a telescope can be considered to be more efficient than another one if more observation tasks can be carried out within the same time interval.
Let us consider that the set of potential tasks is large enough, and we want to select an optimal subset and order it by time.
In order to be able guarantee optimality in the formal sense by minimisation of some cost function, one has to know how to estimate the exposure time and its confidence range.

The required exposure time is determined by a number of parameters that characterise the detecting apparatus, telescope optics, the atmosphere, the sky brightness, and the target object itself.
Given that all these parameters are known within some given precision, calculation of the required exposure time for scheduling observation tasks resolves itself into quite simple operations that can yield the answer with known precision.

The line-of-sight intensity of the optical turbulence (OT) determining the seeing is a stochastic quantity that varies randomly around a typical value over the time.
As a consequence, calculated estimates of the required exposure times also take stochastic form.

Currently, much attention is being given to the development of automatic observation scheduling for telescopes so as to increase the yield of scientific data, and the problem of coming-night forecasting of the OT intensity is acute.
Several different approaches may be used for this.
First, there are techniques based on physical simulations that employ measured meteo parameters~\cite{Masciadri2013I,Giordano2013,Trinquet2006}, and these require some model assumptions.
Second, there are techniques that consider the formal statistical properties of the measured atmospheric turbulence~\cite{Racine1996,Skidmore2009,Kornilov2012}.

We will approach this problem by ignoring the physical origins of the OT intensity variation, which will be formally described as some time series.
Specifically, common simple auto-regressive integrated moving average (ARIMA) models will be used.
Likely for the first time, an ARIMA model was referenced in context to the seeing several decades ago~\cite{Aussem1994}, but it was not applied directly to atmospheric turbulence parameters.
Since then, information about the development of this idea has not been described in the literature, but the current accumulated bulk of real measurements allows us to consider this approach in detail.

The effect of the in-dome turbulence has not been taken into account in this paper because the methods for its elimination are known in general.
At the same time, it is not possible to vanish OT in the atmosphere.

The paper is organised as follows.
In Section~\ref{sec:feat}, the characterisation of the analysed data is presented, and a linear model of the data is proposed in Section~\ref{sec:arima} along with one modification, which is presented in Section~\ref{sec:demean}.
In Section~\ref{sec:fcst}, the forecasting results for the central image intensity, full width at half maximum~(FWHM), and the radius encircling $80\%$ of the energy are presented.
Validation of the forecasting and Monte-Carlo simulations are also carried out.
In Section~\ref{sec:outro}, the results are discussed and the major conclusions are provided.

\section{Data features}
\label{sec:feat}

\subsection{Basic concepts}
Modern ideas of the atmospheric OT~\cite{Tatarskii1967,Roddier1981} assume that when the Kolmogorov turbulence model is used, the structural coefficient of the refractive index $C^2_n$ is the only quantitative parameter needed to characterise the intensity of the OT.
Under the assumption of independence of layers, the integrated line-of-sight effect can be expressed by the OT intensity as follows:
\begin{equation}
J=\int_0^{\infty} C^2_n(h) dh.
\end{equation}

The most important and well-known characteristic is the seeing~$\beta$, which is the FWHM parameter of the atmospheric point spread function (PSF)~\cite{Roddier1981}.
When a Kolmogorov turbulence spectrum is assumed, the seeing is related to the full line-of-sight OT intensity as follows:
\begin{equation}
\label{eq:seeing}
\beta = 5.307 \cdot \lambda^{-1/5} J^{3/5}\,\mbox{radian} \approx 2 \cdot 10^{7} J^{3/5}\,\mbox{arcsec.},
\end{equation}
where $\lambda$, the wave length, is $500\ \mbox{nm.}$~\cite{Tokovinin2007}.
Alternatively, it also can be expressed through well-known Fried $r_0$ parameter~\cite{Fried1966} as follows:
\begin{equation}
\label{eq:r0}
\beta = 0.98 \frac{\lambda}{r_0}.
\end{equation}

In this paper, all possible effects related to distinctions from Kolmogorov turbulence, for instance, the outer scale, are ignored mostly because of the lack of reliable data on these features; however, we hope to account for them in the future if possible.

Let us now briefly recall the major principles involved in recovering the OT strength from multi-aperture scintillation sensor/differential image motion monitor (MASS/DIMM) measurements.
Since a full description is given in~\cite{Tokovinin2007,Kornilov2011c}, the following summary reports only the facts important for the present work.
The observables of the MASS channel are so-called scintillation indices $s^2_{ij}$, where $i,j = 0,1,2,3$ and enumerate the device input apertures of the different sizes.
Optical radiation passes through the turbulent atmosphere and is received via the $i$-th aperture.
The scintillation indices are variances and covariances of the fluctuations of the light flux that is received through the different apertures.
The observables for DIMM are the longitudinal and transversal variances~$\sigma^2_{l,t}$ for the distance separation of the two images of the same star~\cite{Sarazin1990}.
All these quantities allow us to obtain the full OT intensity~$J$ and its crude vertical distribution at a given time moment~\cite{Kornilov2011c}.

\subsection{Automatic seeing monitor}

To develop a model for the evolution of~$J$ over time, real data obtained from automatic seeing monitor (ASM) measurements taken at the Caucasian Mountain Observatory on Mount~Shatdzhatmaz~($42^\circ 40.00^\prime\,\textrm{N}$ $43^\circ 44.20^\prime\,\textrm{E}$) have been used.
The monitor is located at an altitude of 2100~m, and it is near the new 2.5~m telescope built by Lomonosov Moscow State University~(MSU)~\cite{Kornilov2010}.
About 300 thousands minute values were obtained from November 2007 to June 2013.
Thus, time series of observations~$J(t_i)$ and their evolution on a greater-than-minute time scale were considered.

The measurements are available on a non-uniform time grid~$t_i$ with intervals close to 1~minute.
The time grid non-uniformity is caused by interruptions of the observations during the night (for instance, interruptions due to bad weather) as well as by inherent features of the observation program (for instance, interruptions for sky brightness measurements).

Because it would be more convenient to work on a strictly uniform grid, from the point of view of analysis and forecasting, the raw data~$J(t_i)$ were linearly interpolated onto a uniform time grid with 1-minute steps.
The series power spectrum is thus decreased at high frequencies because of this operation~\cite{Schulz1997}.
However, it is the low-frequency range that is of interest to us from the point of view of the forecast.
Furthermore, in regards to online forecasting, present atmospheric data being obtained by MASS/DIMM in real time will be also interpolated.
In this way, the stochastic time series obtained by interpolating onto a uniform time grid were used instead of the initial raw time series.

\subsection{Transformation of the one-point density probability function}

\begin{figure}[t]
\begin{tabular}{ll}
\includegraphics[height=2.2in]{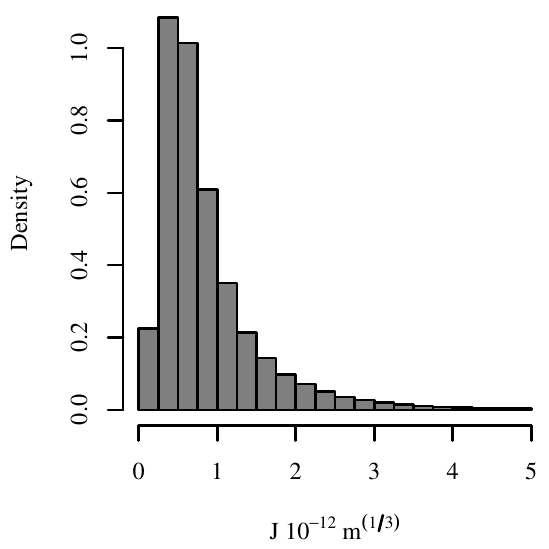} &
\includegraphics[height=2.2in]{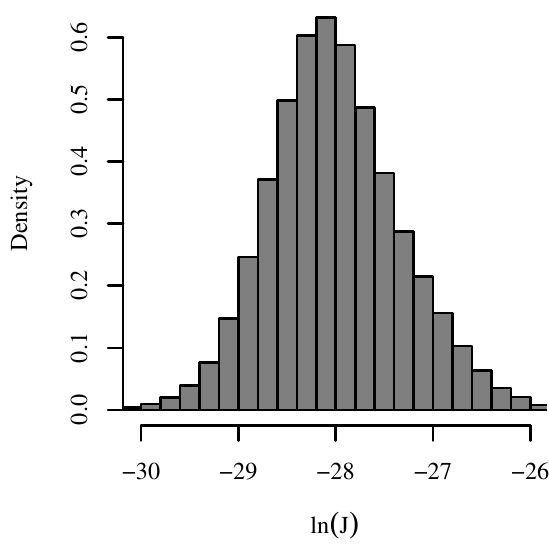}
\end{tabular}
\caption{\label{fig:feat:hist} Left: histogram of the OT intensity $J$ compiled from the data obtained from ASM measurements. The median is $6.6 \cdot 10^{-13}\,\mbox{m}^{1/3}$.
Right: histogram of $\ln\left(J\right)$. The mean is $-28.01$, the standard deviation is $0.673$, and the median is $-28.05$.}
\end{figure}

The one-point probability density function (PDF) is an important characteristic describing the stochastic process.
Using a sampling of the available data, let us plot a histogram of $J$ quantities as an estimate of the one-point PDF.
Please note that the histogram is an estimate of the averaged-over-time PDF when the true probability density has an explicit dependency on time.
This issue is addressed in Section~\ref{sec:feat:time}.
The seasonal behaviour of the data investigated below has been already corrected by the means discussed in the mentioned section.

The histogram of $J$ is shown on the left part of Fig.~\ref{fig:feat:hist}.
One can see that the distribution is asymmetrical.
The histogram of $\ln\left(J\right)$ is depicted on the right part of Fig.~\ref{fig:feat:hist}, and this distribution visually corresponds to a normal one.

The considered quantities have to be normally distributed from the point of view of the techniques applied in Section~\ref{sec:arima}.
Consequently, it is necessary to choose a functional transformation $f$ for the quantities of $J$ that will result in the quantities of $f(J)$ being normally distributed.
Let us note that in these circumstances, the choice of the $f$ function defines the model distribution of the quantity $J$ itself at once, and the quantities for $J$ will be part of more complex expressions and variables in Section~\ref{sec:fcst}; therefore, it would be desirable that there exists an analytic expression of the $J$ moments.

The following well-known broad class of Box--Cox transformations~\cite{Box1964} is usually applied to solve this problem:
\begin{equation}
\label{eq:bc}
f(J,\lambda) =
\begin{cases}
\dfrac{J^\lambda - 1}{\lambda} & \text{if } \lambda \neq 0,\\
\ln{(J)} & \text{if } \lambda = 0.
\end{cases}
\end{equation}
The logarithmic transformation represents a special case, whereby this is the only one special case that allows for the calculation of the moments not in the form of series, but as elementary functions~\cite{Freeman2006}.
The transformation that will be optimal for the data by Box--Cox criterion (maximum likelihood) is the one with an index of power $\lambda=-0.085$, and despite the apparent proximity to the logarithmic transformation, the difference is statistically significant.

\begin{figure}[t]
\begin{minipage}[t]{0.48\textwidth}
	\includegraphics[height=2.2in]{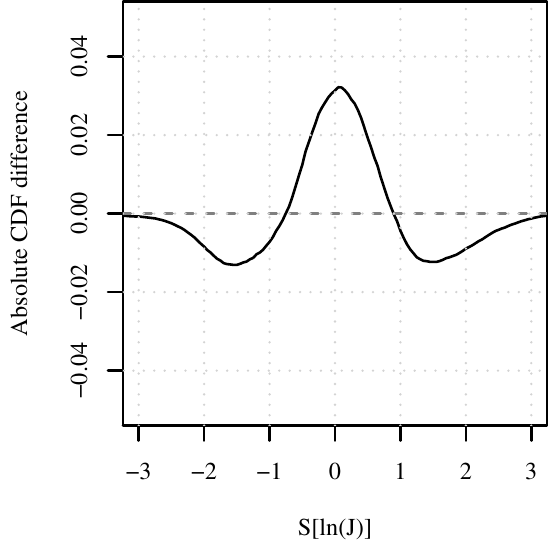}
	\caption{\label{fig:feat:qq} Difference between the cumulative distribution function (CDF) of the quantity $S\left[\ln\left(J\right)\right]$ and the normal distribution with zero mean and unit variance. Here, $S\left[x\right]\equiv\frac{x- \mean{x}}{\sqrt{\var{x}}}$.}
\end{minipage}
\hspace{0.05cm}
\begin{minipage}[t]{0.48\textwidth}
	\includegraphics[height=2.2in]{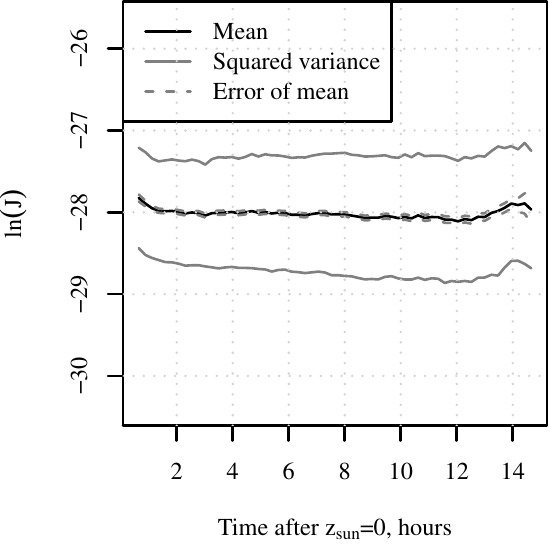} \\
	\caption{\label{fig:feat:nightly} Mean and standard deviation of the data versus time from the sunset~(Sun altitude $z_\text{sun}=0$).}
\end{minipage}
\end{figure}

Therefore, the simple logarithmic transformation $x(t_i) = \ln J(t_i)$ has been chosen.
The difference between the cumulative distribution function of the empirical distribution of the $x$ values and the theoretical normal distribution is given in Fig.~\ref{fig:feat:qq}, where $\mean{\cdot}$ and $\var{\cdot}$ denote operations of the mean and the variance, respectively.
One can see that the approximation error does not exceed $0.03$.
The similarity between the log-normal distribution and the distribution of the seeing has already been noted many times before, for instance, by Racine~\cite{Racine1996}.
In this assumption, both $\beta$ and $r_0$ are log-normally distributed because of relations~(\ref{eq:seeing}--\ref{eq:r0}).

A question arises here as to whether the chosen distribution approximation is good enough.
The distribution approximation error could lead to, for instance, an unacceptable bias in the estimations and to an underestimation of the variance.
The answer to this question will be given in Section~\ref{fcst:validation}, where cross-validation of the whole forecasting procedure is carried out.

\subsection{On the deterministic dependency on time}
\label{sec:feat:time}

The question about the character of dependency of the analysed quantities on time is an important one.
Is it possible to extract such an additive component that explicitly depends on time from the data series so that the residual probability density does not depend on time?

There are two types of expected deterministic behaviours for the OT strength; there are the daily behaviour related to sunset and related physical phenomena in the Earth's atmosphere, and the annual one related to the change of the seasons.

The dependency of the mean and standard deviation of $\ln J(\tau)$ on time $\tau$ from the sunset is presented in Fig.~\ref{fig:feat:nightly}.
One can see an initial trend at the first hour after the sunset.
More than 1.5~hours are always required for the Sun to fall between $0$ and $18$ degrees under the horizon at our site; thus the indicated time is not an astronomical night.
Therefore, we can say that there is no dependence of the mean and the variance of the distribution on time within the observational night.
Behaviour after 12 hours was caused by small amount of long nights which occur only during the winter season.
This feature was noted in an earlier paper~\cite{Safonov2011}.

\begin{figure}[t]
\centering
\begin{tabular}{c}
\includegraphics[width=4.4in,height=3.6in]{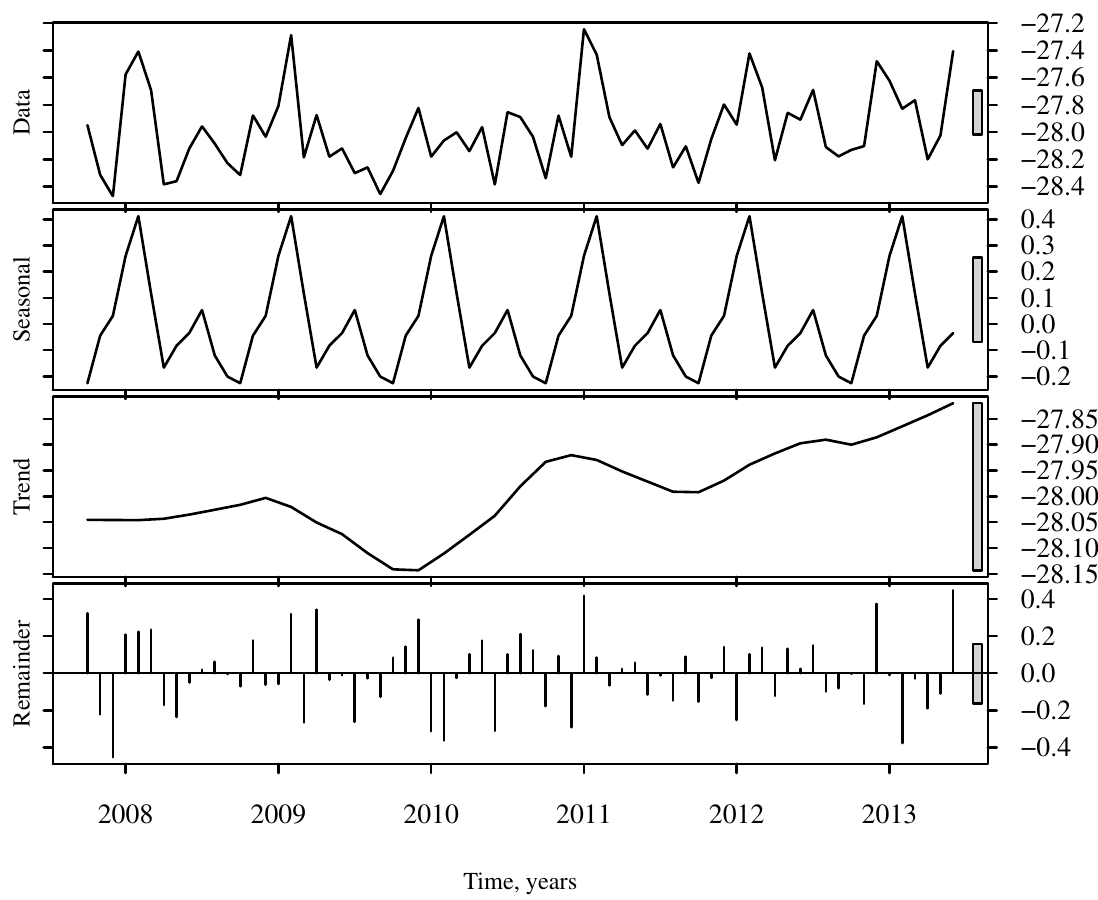}
\end{tabular}
\caption{\label{fig:feat:stl}Decomposition of the monthly averaged logarithm of OT intensity into the seasonal, trend and irregular parts. Gray rectangles at the right side, which have the same height, are to show the relative scale.}
\end{figure}

Calculations of the decomposition of monthly averaged values of $\ln\left(J\right)$ into a seasonal, trend and irregular parts were accomplished by the technique {\tt stl}~\cite{Cleveland1990} of the standard package {\tt stats} of analytic system R~\cite{R2013}; these data are presented in Fig.~\ref{fig:feat:stl}.
The annual seasonal behaviour is notable, and it has been mentioned, for instance, in paper~\cite{Kornilov2010}.
However, the magnitude of the effect~(the amplitude of the seasonal variations reached $0.6$) was comparable in value to the the standard deviation of the irregular component, which was $0.2$.
The inference that the greatest variation is observed between nights but not months was confirmed in an earlier paper~\cite{Kornilov2014}.
The given seasonal behaviour is easily allowed by additive correction of the source data.

Thereby, deterministic behaviour will not being considered further in the present paper because the stochastic one dominates.
However, the forecasting approach being used in this paper does not exclude the possibility of considering additive deterministic features.
In the case of annual dependency, these features have been taken into account.
The deviations of monthly averaged values from their means were subtracted from source data, and the standard deviation decreased by $4\%$.

\section{Linear auto-regressive moving average models}
\subsection{Conventional model}
\label{sec:arima}

The linear stochastic auto-regressive moving average series $x_i$ is defined by recursive differential equations as follows:\footnote{As a result of the linearity, $\mean{x_i}=0$ in this section without loss of generality.}
\begin{equation}
x_{i} - \sum_{j=1}^{p}x_{i-j} \phi_j = a_{i} + \sum_{j=1}^{q} \theta_{j} a_{i-j},
\label{eq:arima:1}
\end{equation}
where $a_{i}$ are independent normally distributed quantities with zero mean and finite variance, and $\phi_j$ and $\theta_k$ are the corresponding $p$ and $q$ real parameters characterising the model denoted as \arima{p}{0}{q}.
There is an exhaustive presentation of the auto-regression theory in the Box and Jenkins monograph~\cite{Box1976}.
Let us now recall some important details.

By `forecasting', we mean the process of calculating the multidimensional conditional PDF (or its parameters) $p(x_l,...x_{l+N} | \hat x_1,...,\hat x_r)$, where $\hat x_i$ are realisations of the investigated process, which are observed at the $i$-th time moment in the past, and $x_l$ are random quantities being forecasted.
The quantities of $a_{i}$ (and thus $x_{i}$) are considered to be normally distributed; therefore, the conditional PDF can be fully defined by the mean and covariance matrix.

The standard package {\tt stats} of the system R~\cite{R2013} has been used to estimate the model parameters.
Given $p$ and $q$, the parameters amounts, the model parameters $\phi_j$ and $\theta_k$ can be calculated by maximising the likelihood function for the presented realisation of the investigated stochastic process.

The observational data for the year~2009 consists of about 45 thousand 1-minute $\ln\left(J\right)$ values interpolated onto a uniform time grid, and these data were used to identify the model (i.e. to determine the values of the $\phi_j$ and $\theta_k$ parameters).
The 1-year subset was used for the following reasons.
First, the data over the whole range displayed some degree of homogeneity, and the use of more than several thousand values to maximise the likelihood was too computationally expensive.
Second, we also needed some unused data to validate the model.
The missed values (the values corresponding to interrupts between subsequent observations of more than 90~seconds) have been substituted by service {\tt NA} marks in the R system, and the standard package {\tt stats} is able to correctly interpret these missed values during parameter fitting.

Akaike informational criterion~(AIC)\footnote{Inherently, the likelihood value corrected by the number of parameters.}~\cite{Akaike1974}, the residual auto-correlation function~(ACF), and the principle of the least number of model parameters are all important criteria for model identification.

\begin{figure}[t]
\centering
\begin{tabular}{cc}
\includegraphics[height=2.2in]{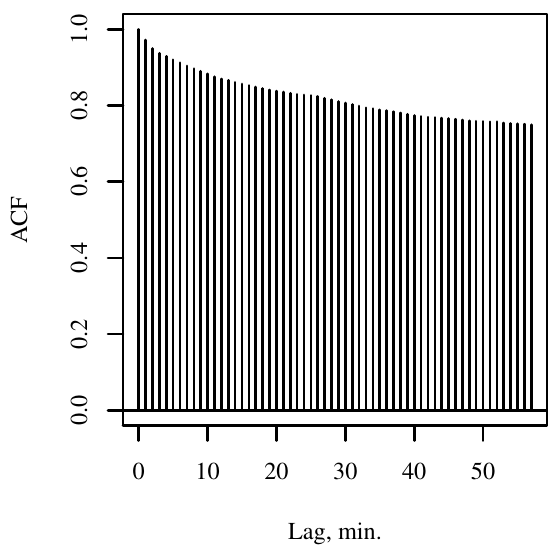}
\includegraphics[height=2.2in]{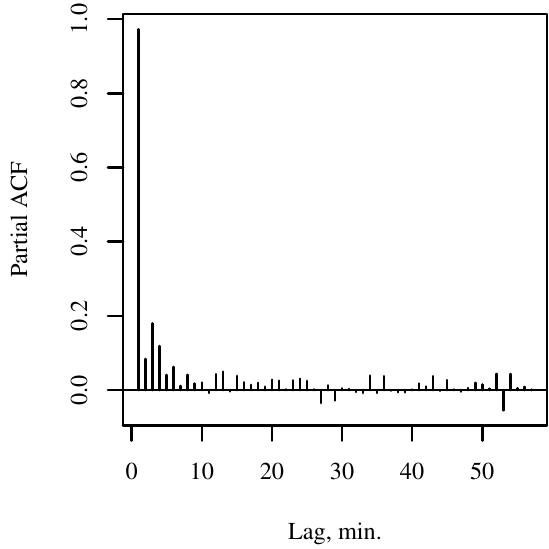}
\end{tabular}
\caption{\label{fig:arima:acf}Left: ACF of the 2009-year data linearly interpolated onto a uniform time grid with 1-minute steps.
Right: Partial ACF of the 2009-year data linearly interpolated onto a uniform time grid with 1-minute steps.
The 95\% confidence interval of the sampling correlation coefficient is well below the plot scale and not shown.}
\end{figure}

The ACF and partial ACF~(see formula~(3.2.33) from~\cite{Box1976} for its definition)~of the analysed data are given in Fig.~\ref{fig:arima:acf}.
It can be confused with ACF of the non-stationary stochastic process, assuming that the process requires the use of the first-order difference or high-order ones.
Based on perceptions of the investigated process, one must require a finite forecast variance at infinity.
This requirement is imposed because the sampling (time-averaged) quantities distribution has a finite variance that must bind the forecast variance at infinity.
This requirement leads to the absence of the unit-roots in the auto-regression part of the model.
From the point of view of model identification, this requirement forces us to abandon models that employ the difference (integrated mean auto-regressive models).

\begin{table}[!tbp]
\caption{Features of some models.
The ACFs of the residuals are given in Fig.~\ref{fig:arima:models}.
In the table here are AIC~\cite{Akaike1974},
the model parameters $\phi_i$ and $\theta_j$ corresponding to equation (\ref{eq:arima:1}),
variance $\sigma^2_a$ of $a_i$ from equation (\ref{eq:arima:1}),
quantities $\rho$ and $T$ which are constants of the asymptotic behaviour of the eventual function,
and number $r$ of the series elements $|\pi_k| > 3\cdot10^{-3}$. \label{table:arima}}
\begin{center}
\begin{tabular}{|l||c|c|c|}
\hline
Model & \arima{3}{0}{4} & \arima{1}{0}{5} & \arima{4}{0}{1} \\
\hline
\hline
AIC & $-50751.49$ & $-50555.73$ & $-50744.43$  \\
\hline
$\phi_i$ & $\phi_1 = 0.84 \pm 0.01$ & $\phi_1 = 0.9956 \pm 0.0003$ & $\phi_1 = 1.806 \pm 0.008$ \\
 & $\phi_2 = 0.9980 \pm 0.0003$ & & $\phi_2 = -0.940 \pm 0.012$ \\
 & $\phi_3 = -0.84 \pm 0.01$ & & $\phi_3 = 0.232 \pm 0.010$ \\
 & & & $\phi_4 = -0.098 \pm 0.006$ \\
\hline
$\theta_j$ & $\theta_1 = 0.04 \pm 0.01$ & $\theta_1 = 0.106 \pm 0.005$ & $\theta_1 = 0.916 \pm 0.007$ \\
& $\theta_2 = -1.08 \pm 0.01$ & $\theta_2 = 0.221 \pm 0.005$ &  \\
& $\theta_3 = 0.009 \pm 0.007$ & $\theta_3 = 0.078 \pm 0.005$ & \\
& $\theta_4 = 0.138 \pm 0.007$ & $\theta_4 = 0.054 \pm 0.005$ & \\
& & $\theta_5 = 0.038 \pm 0.005$ & \\
\hline
$\rho$ & 0.997 & 0.996 & 0.998 \\
$T$ & $\approx 425$ & $\approx 230$ & $\approx 490$ \\
$\sigma_a^2$ & 0.018 & 0.018 & 0.018 \\
$r$ & 15 & 12 & 16 \\
\hline
\end{tabular}
\end{center}
\end{table}

\begin{figure}[htp]
\begin{center}
\includegraphics[width=4.6in]{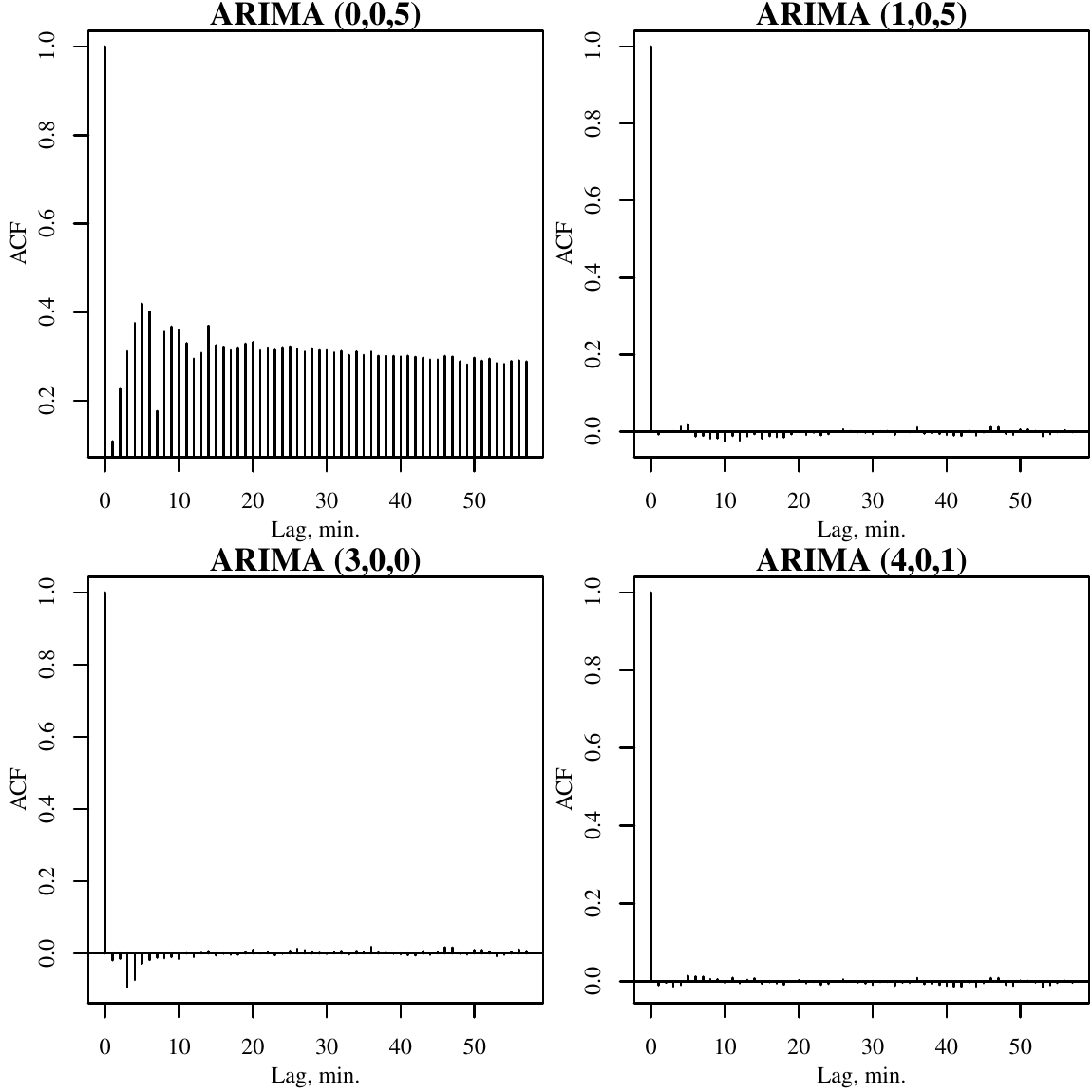}
\caption{Auto-correlation function of the residuals of some models.
The 95\% confidence interval of the sampling correlation coefficient is well below the plot scale and not shown.
One can see that not all models describe the correlative data properties well.
In the left column, the models with considerable residual correlations are presented.\label{fig:arima:models}}
\end{center}
\end{figure}

The parameters of the models with parameter numbers of $p \le 5, q \le 5$ have been found.
Moreover, due to the fact that the problem can be reduced to finding the quadratic function minimum and requires less computational resources than in the case of pure auto-regression ($q=0$), the parameters for considerable numbers of models with $q=0, p \le 20$ have been also found.
All the models can relatively be grouped into the following two different classes: ones leading to uncorrelated residuals\footnote{That is one of the criteria for model identification adequateness.} and ones with considerable residual correlations.
The models of the first class demonstrate similar behaviour for the eventual function~(conditional mean $\mean{x_l|\hat x_1,...,\hat x_r}$ as a function of $l$) and the variance with large $l$ ($l > 10$), and the forecast mean exponentially tends to the process mean while the variance tends to the unconditional process variance.
Several examples of such models are given in Fig.~\ref{fig:arima:models} and in Table~\ref{table:arima}, where the asymptotic behaviour of the eventual function is approximated by the exponential function $A \rho^l + m = A \exp(-\frac{l}{T}) + m$.

From the practical viewpoint of forecasting different quantities, the particular type of model is of no importance because all expressions for the forecast mean, variance, and covariance calculations are the same for all linear auto-regressive moving average models.
Here, the model \arima{4}{0}{1} will be used for demonstration and simulation purposes (see the right panel of Fig.~\ref{fig:arima:models}).

The forecast calculation procedure is reduced to the calculation of the parameters of the conditional multidimensional PDF.
This is usually done as follows.
\begin{itemize}
\item{
The conditional means $\mean{x_l|\hat x_1,...\hat x_r}$ are calculated via model parameters and initial values by means of equation~(\ref{eq:arima:1}); then it is possible to solve the equation for $x_i$ as follows:
\begin{equation}
\label{eq:arima:pi}
x_{i} = \sum_{k=1}^{\infty} \pi_{k} x_{i-k},
\end{equation}
where $\pi_k$ are the weights expressed through $\phi_j$ and $\theta_k$~({see formulae~(5.2.3) and (A5.2.1) from~\cite{Box1976}}).
The number $r$ of required initial values is determined based on the values of the weights $\pi_k$.
Only the first $p$ weights of $\pi_k$ are always distinct from zero for pure auto-regression models ($q=0$)~\cite{Box1976}.
}
\item{
The covariance matrix is fully determined by the model parameters $\phi_j$ and $\theta_k$, and it does not depend on initial conditions~({see formulae~(5.2.3) and (A5.1.1) from~\cite{Box1976}}).
}


\end{itemize}

The model performance will be illustrated later in Section~\ref{sec:fcst}.
Let us now dwell on a simple obvious model modification, which was inspired by the well-known work of Racine~\cite{Racine1996}.

\subsection{On the average per night: the modified model}
\label{sec:demean}

\begin{figure}[t]
\centering
\includegraphics[height=2.2in]{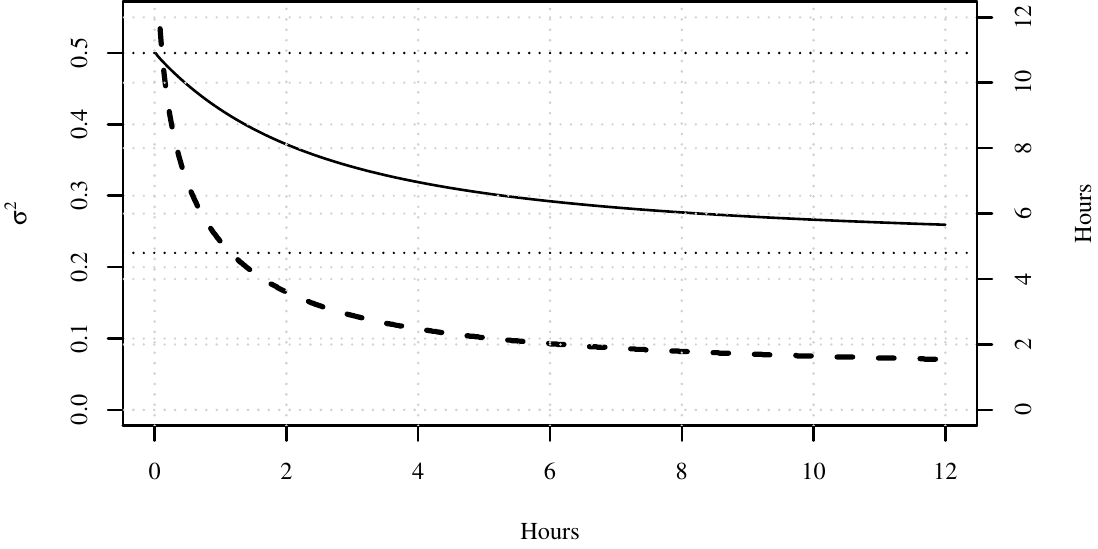}
\caption{\label{fig:demean} 
Limiting variance $\breve{\sigma}^2_{x,\infty}$ of the forecast of the model with the constant night mean versus number of measurements used to estimate it.
The dashed line corresponding to the right axis shows the time interval after which the modified model performed better forecasts than the original one versus the number of measurements used to estimate mean over the night.}
\end{figure}

The obvious desire was to improve the model in the sense of decreasing the variances of the forecasted quantities.
The variation of the OT strength inside a single night is usually less than the variation of the mean for different nights.
As such, it is seems natural to consider the following modified model for the forecasted quantity behaviour:
\begin{equation}
x_{i} = z_{i} + m_{j(i)},
\label{eq:demean:1}
\end{equation}
where $x_{i}$ are the considered quantities, $m_{j(i)}$ is the stochastic quantity of the level average over night, and $z_i$ are deviations from the mean, $\mean{z_i}=0$.
This model assumes that for any measurement $i$ related to the night $j$, there is a particular value for the average level $m_{j(i)}$ that persists as a constant over the single night, but varies from one night to another.
Let us now get rid of index $j(i)$ because only the single night forecast is of interest.

In the current model, $x_{i}$ are still the observables, but one can a posteriori calculate $z_{i}$ and construct the auto-regressive moving average model for these values in a manner similar to the techniques described in the previous sections.
Indeed, unconditional (limiting) variance of $z_{i}$ appears to be almost half as less than that for $x_{i}$.
However, during the online forecasting process, it is not possible to know the exact value of the $m$ realisation, only its estimate.
The estimate is required for use with expression~(\ref{eq:demean:1}) in order to convert from $\hat x_i$ to $\hat z_i$ and to backwardly convert the forecast from $z_l$ to $x_l$.
One can show that resulting forecast variance is
\begin{equation}
\breve{\sigma}^2_{x,l} = \sigma^2_{z,l} + \sigma^2_m \left(1-\sum_{j=1}^{r}\psi_{r+l-j}\left(1 - \sum_{k=1}^{j-1} \pi_{j-k} \right) \right)^2.
\label{eq:demean:2}
\end{equation}

Moreover, the maximum possible unconditional variance $\breve{\sigma}^2_{x,\infty}$ must coincide with the unconditional variance of the unmodified model $\sigma^2_{\infty}$.
The analysis of the 2009-year data shows that the unconditional variance $ \sigma^2_{x,\infty} \approx 0.5 $ and $\sigma^2_{z,\infty} \approx 0.22$.
Hence, the unconditional variance $\sigma^2_m \approx 0.28$.

There are two natural ways to construct the estimate for the quantity $m$.
The first method uses the average values of the past nights.
However, data show that series $m_j$, which consist of average-over-night values, represent an auto-regression process of the first-order with coefficient $\phi_{1} \approx 0.5$.
Hence, the conditional variance $m_{j+1}$ with known $\hat m_j$ is only 75\% of the unconditional variance value.

The second method is potentially more efficient and is based on usage of the mean of the values measured in the current night as an estimate of $m_j$.
As more and more measured values for $\hat x_i$ are received, the variance of such a quantity will decreases in the following way:
\begin{equation}
\sigma^2_{m} \sim \frac{1}{N} + \frac{2}{N^2} \frac{\rho}{1-\rho} \left(N - \frac{1-\rho^N}{1-\rho} \right),
\label{eq:demean:3}
\end{equation}
where $\rho$ is the correlation coefficient between two consecutive quantities $x_{i}$ and $N$ is the number of measurements used to estimate $m$.
Due to the fact that in reality $\rho \to 1$, $\sigma^2_m$ reduces down almost linearly with increasing $N$.

The dependence of the limiting variance $\breve{\sigma}^2_{x,\infty}$ on the quantity $N$ is shown in Fig.~\ref{fig:demean} with respect to equations~(\ref{eq:demean:2})~and~(\ref{eq:demean:3}).
Let us compare the dependence of the forecast variance values $\breve{\sigma}^2_l$ of the modified model and  $\sigma^2_l$ of the unmodified one on forecasting time moment $l$.
For any $\sigma^2_m$, the forecast variance of the modified model $\breve{\sigma}^2_l$ grows with increasing $l$ faster than $\sigma^2_l$ for small $l$.
But, it is majorised by the values from $0.22$ to $0.5$ depending on $\sigma^2_m$.

Accordingly, for any $\sigma^2_m$ (that in turn depend on $N$), there is such a time moment $l^{\ast}$ that $\breve{\sigma}^2_{x,l^{\ast}} = \sigma^2_{x,l^{\ast}}$.
For $l < l^{\ast}$, $\breve{\sigma}^2_{x,l} > \sigma^2_{x,l}$.

For data from the year~2009, the dependence of $l^{\ast}$ on quantity $N$ is presented in Fig.~\ref{fig:demean} by the dashed line.
For all practical values of $N$, the lower limit of quantity $l^{\ast}$ was two hours.
The use-case of the modified model is quite limited by cases at the middle and end of the continuous nights.
Moreover, one always has to sacrifice the forecast quality for $l < l^{\ast}$.

The question of which time range is more important remains open here.
One cannot approach this problem from the point of view of variance comparisons anymore, thus additional criteria must be involved.
For instance, it is obvious that the probability of bad weather appearing is higher with large $l$; therefore the probability that the forecast will be unused is higher.

In what follows, we use the conventional model from Section~\ref{sec:arima}.

\section{Forecast}
\label{sec:fcst}

\subsection{Image parameters being forecasted}

What matters most is not the seeing itself but integrated-over-exposure-time image characteristics in real astronomical tasks related to obtaining images.
As such, we would like to obtain statistical properties for the stochastic quantities derived from the forecast.
Three different parameters characterising long-exposure images are considered further here.
These are as follows.
\begin{itemize}
\item{The central intensity of the image for unit flux with a given exposure time in the focal plane of a large ideal telescope.
This value is proportional to the Strehl ratio.}
\item{The FWHM of the PSF for cases requiring an angular resolution.}
\item{Angular size of the PSF containing fraction $e$ of the full energy for cases requiring contrast achievement.}
\end{itemize}

To begin, let us set the functional relations connecting these parameters that we are interested in, with 1-minute instant seeing values.
The single 1-minute PSF is considered to follow the two-dimensional Gauss function, where the current seeing is its FWHM parameter~\cite{Roddier1981}.
The detector is considered to be ideally linear, then the expression for the central intensity $\gamma_1$ is obtained as follows:
\begin{equation}
\label{eq:central_int}
\gamma_1 = \frac{4 \ln 2}{\pi N}\sum_{i=1}^{N} \frac{1}{\beta^2_i},
\end{equation}
where N is the number of minutes in the exposure time and $\beta_i$ are the corresponding instant seeing random values.
The value of $\gamma_1$ is measured in units of inverse squared arcseconds.

The FWHM of the integrated-over-time PSF represented by the two-dimensional Gaussian function with changing-in-time parameters has no simple analytic expression.
Instead, this quantity $\gamma_2$ is defined by the following algebraic equation:
\begin{equation}
\label{eq:fwhm}
F_2(\gamma_2, \beta_1, ..., \beta_N) \equiv \sum_{i=1}^{N}\frac{1}{\beta^2_i}\left(\exp\left(-\frac{ \gamma_2^2 \ln 2}{\beta^2_i}\right)-\frac{1}{2}\right) = 0.
\end{equation}

Similarly, the circle size (radius) $\gamma_3(e)$ with the given amount of energy is defined as follows:
\begin{equation}
\label{eq:energy_circle}
F_3(\gamma_3(e), \beta_1, ..., \beta_N) \equiv 1 - e - \frac{1}{N}\sum_{i=1}^{N}\exp\left(-\frac{4 \gamma^2_3(e) \ln 2}{\beta^2_i}\right) = 0.
\end{equation}

The notation in expressions~(\ref{eq:energy_circle})~and~(\ref{eq:fwhm}) corresponds to that it expression~(\ref{eq:central_int}).
Further, for the sake of simplicity, $\gamma_3$ without an argument denotes just $\gamma_3(0.8)$.

It is of interest that the type of dependency of $\gamma_2$ on $\beta_i$ is such that decreases in the instant seeing $\beta_i$ values improve the results more than increases in equal-in-values worsens the results.
Thus, the typical FWHM value obtained with an exposure time of several minutes is a little narrower than the mean instant seeing value.

\subsection{Forecasting}
\label{fcst:build}

The relation between the seeing $\beta_i$ expressed in arcseconds and the full line-of-sight OT intensity is given by formula~(\ref{eq:seeing}), where $\ln\left(J_i\right) = x_i$.
As soon as $x_i$ are normally distributed, then both $J_i$ and $\beta_i$ follow log-normal distributions in turn.
By designating the means and covariances of joint distribution $x_i$ by $\mu_i$ and $\sigma_{ij}$ correspondingly, the following relation can be established:
\begin{equation}
\label{eq:2}
\mean{\ln \beta_i} \equiv \hat \mu_i = \frac{3}{5} \mu_i + 7 \ln 10 + \ln 2,
\end{equation}
\begin{equation}
\label{eq:3}
\cov{\ln \beta_i}{\ln \beta_j} \equiv \hat \sigma_{ij} = \left(\frac{3}{5}\right)^{2} \sigma_{ij},
\end{equation}
where $\mean{\cdot}$ still designates the mean and $\var{\cdot}$ designates either the variance or the covariance of two random variables.
The means and covariances of $\beta_i$ are expressed through those of $\ln\left(\beta_i\right)$ by formulae for calculation of the log-normal distribution moments.
If $\mu_i$, $\sigma_{ij}$ are means and covariances of the normal distribution, and $m_i$, $s_{ij}$ are corresponding moments of the log-normal one, then
\begin{equation}
\label{eq:3a}
m_i = \exp\left(\mu_i+\frac{1}{2}\sigma_{ii}\right),
\end{equation}
\begin{equation}
\label{eq:3b}
s_{ij} = \exp\left(\mu_i + \mu_j + \frac{1}{2}\sigma_{ii}+ \frac{1}{2}\sigma_{jj}\right)\left(\exp\left(\sigma_{ij}\right) - 1\right).
\end{equation}

While obtaining convenient analytic expressions for PDFs of $\gamma_j$ would be the most desirable, this is likely not possible.
Because of this, let us find an approximation of the distributions by the technique put forward by Fenton~\cite{Fenton1960}.
In that work, a simple technique to approximate the sum of log-normal quantities by the log-normal distribution was proposed.
It is important to note that the quantities being summed are not correlated in the earlier paper~\cite{Fenton1960}, but our quantities are correlated.
Moreover, equations~(\ref{eq:energy_circle})~and~(\ref{eq:fwhm}) have the form of the sum of the log-normally distributed quantities only when Taylor series decomposition has been applied in the following way:
\begin{equation}
\gamma_j(\varrho_1, ..., \varrho_N) \approx
\gamma_j(\mean{\varrho_1}, ..., \mean{\varrho_N}) +
\sum_{i=1}^{N} \left.\frac{\partial \gamma_j}{\partial \varrho_i}\right|_{ \varrho_k=\mean{\varrho_k}  } \left( \varrho_i - \mean{\varrho_i} \right),
\end{equation}
where the notation $\varrho_i \equiv \frac{1}{\beta_i^2}$ used here has been introduced for brevity and $\varrho_k$ represent $N$ arguments of the partial derivative.

One may doubt the applicability of the proposed technique.
However, the approximation is eventually compared with the distribution obtained by some kind of simulation both in the previous paper~\cite{Fenton1960} and in the present one~(see Section~\ref{fcst:validation}).
In this manner, the applicability of the technique for correlated and initially not quite log-normal quantities is not being proven here,
but the mean and the variance of $\gamma_j$ can be found in the linear approximation (it is exact for $\gamma_1$) as follows:
\begin{equation}
\label{eq:approx_mean_gamma_j}
\mean{\gamma_j( \varrho_1, ..., \varrho_N)} \approx \gamma_j(\mean{\varrho_1}, ..., \mean{\varrho_N}),
\end{equation}
\begin{equation}
\label{eq:approx_var_gamma_j}
\var{\gamma_j( \varrho_1, ..., \varrho_N)} \approx \sum_{i,k=1}^{N}  \left.\left(\frac{\partial \gamma_j}{\partial \varrho_i}   \frac{\partial \gamma_j}{\partial \varrho_k} \right)\right|_{\varrho_{m,l}=\mean{\varrho_{m,l}}} \cov{\varrho_i}{\varrho_k}.
\end{equation}
The partial derivatives of $\gamma_{2,3}$ are calculated with respect to the implicit function theorem and using equations~(\ref{eq:fwhm}) and (\ref{eq:energy_circle}).

The mean and the covariance of the log-normally distributed quantities $\varrho_i$ are exactly given by expressions~(\ref{eq:3a}) and (\ref{eq:3b}).

The last task that remains is to find the parameters of such a log-normal distribution that would have the given mean~(\ref{eq:approx_mean_gamma_j}) and the given variance~(\ref{eq:approx_var_gamma_j}) as described in the earlier paper~\cite{Fenton1960}.
Accordingly, the approximating distribution of the quantities  $\gamma_i$ is fully defined.

\begin{figure}[t]
\begin{minipage}[t]{0.49\textwidth}
	\includegraphics[height=2.2in]{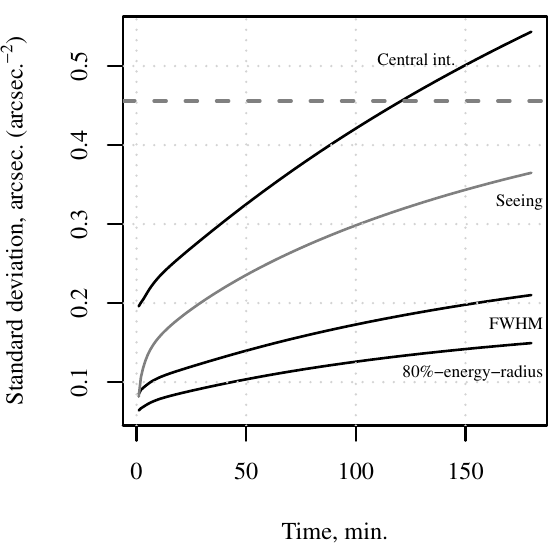}
	\caption{\label{fig:fcst:variance} Averaged (by initial values) conditional standard deviations of the PSF central intensity~($\gamma_1$), PSF FWHM~($\gamma_2$), the radius encircling 80\% of the energy ($\gamma_3$), and the 1-minute seeing values~($\beta$) versus time. All the values are expressed in arc seconds, except the central intensity that is expressed in inverse squared arc seconds. The dashed line shows the unconditional seeing standard deviation of $0.46\,\mbox{arcsec}$.}
\end{minipage}
\hspace{0.05cm}
\begin{minipage}[t]{0.49\textwidth}
	\includegraphics[height=2.2in]{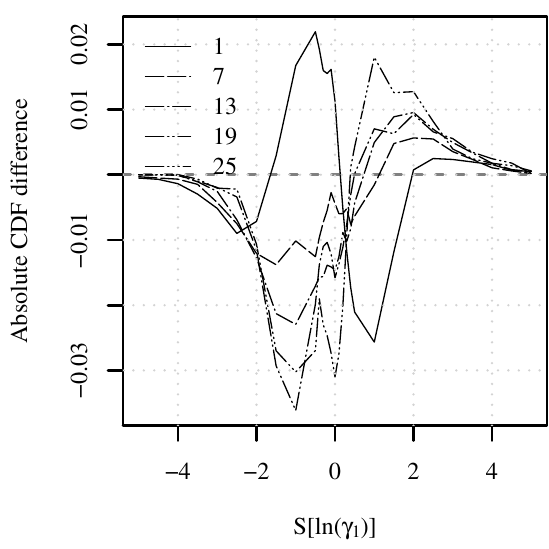}
	\caption{\label{fig:fcst:ce} Difference between the normal distribution with zero mean and unit variance and the cumulative distribution function (CDF) $F\left(\ln\left(\gamma_1\right)\right)$ of the logarithm of the PSF central intensity. Different lines show lags from 1 to 25 minutes.
}
\end{minipage}
\end{figure}

The averaged-over-initial-values conditional standard deviations of the considered quantities $\gamma_j$ are given in Fig.~\ref{fig:fcst:variance}.
The plots demonstrate the informational content of the forecast, and one can see that the process does not fully `forgets' its initial state within about 3~hours, although the standard deviation of the seeing tends to the unconditional one.
Note that there is a property of ARIMA whereby the forecast variance of $\ln\left(J_l\right)$ (and $\ln\left(\beta_l\right)$) depends only on the time advance $l$ and not on initial values.
However, the variance of $\beta$ itself is calculated using~(\ref{eq:3b}) and thus depends on initial values.
One should therefore expect greater variances for greater mean values for all forecasted quantities.

To summarise, let us recall the order of the $\beta_i$ and $\gamma_j$ distribution construction:
\begin{itemize}
\item the parameters of the conditional covariance matrix $\sigma_{ij}$ of $x_l$ are calculated by means of the parameters $\phi_i$ and $\theta_j$;
\item the conditional means $\mu_i$ are calculated using the initial values $\hat x_i$ and the parameters $\phi_i$ and $\theta_j$;
\item the parameters of the joint distribution of $\beta_i$ are calculated with respect to ~(\ref{eq:2})--(\ref{eq:3b});
\item for the given $N$, the means and the variances (that can be converted to log-normal distribution parameters by (\ref{eq:3a}) and (\ref{eq:3b})) of $\gamma_j$ are calculated using (\ref{eq:approx_mean_gamma_j}) and (\ref{eq:approx_var_gamma_j}).
\end{itemize}

\subsection{Forecast validation}
\subsubsection{The Monte-Carlo approach}
\label{fcst:validation}
In order to check the correspondence between the distributions of the real quantities $\gamma_j$ and the model ones, a Monte-Carlo approach will be used.
For this, we use observation data from the year~2010 (the model has not been trained on this data), which is similar in terms of its characteristics and size to the considered 2009-year data.

The time moment was chosen randomly, the forecast was carried out, and the same quantities were calculated from the known realisation.
As soon as the distribution of $\gamma_j$ was considered to be close to log-normal~(see Section~\ref{fcst:build}), we compared quantities $\ln \left(\gamma_j\right)$, and their distribution was close to normal in this case.
Assuming that the cumulative distribution function is $F(\ln \gamma_j) \equiv \Phi(\frac{\ln \gamma_j - \mu}{\sigma})$ for the same time moment, the cumulative distribution function $\Phi$ can be recovered from our experimental sample because $\mu \equiv \mean{\ln \gamma_j}$ and $\sigma^2 \equiv \var{\ln \gamma_j}$ were determined during forecasting.

\begin{figure}[t]
\centering
\begin{tabular}{cc}
\includegraphics[height=2.2in]{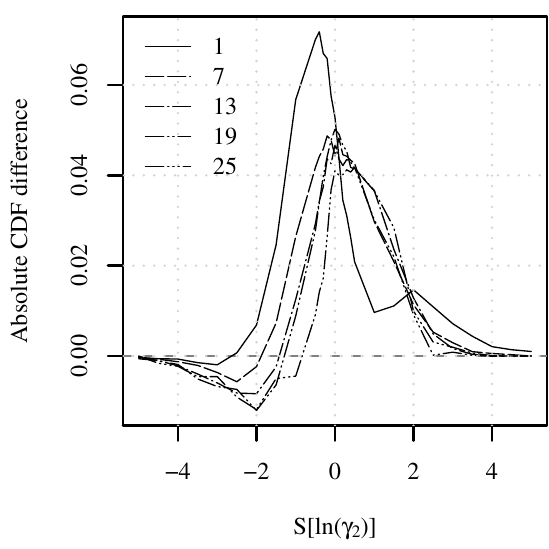} &
\includegraphics[height=2.2in]{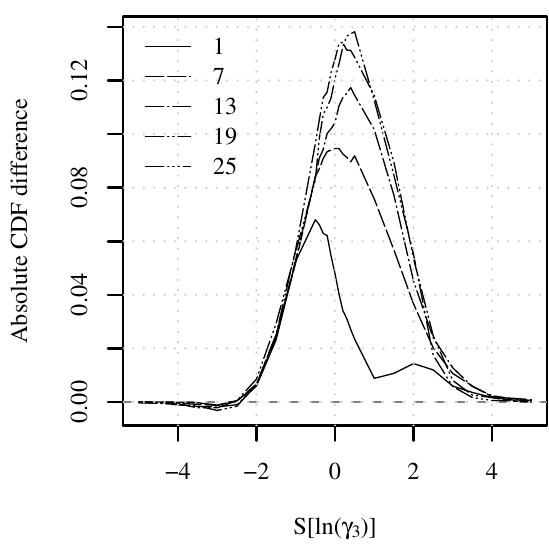}
\end{tabular}
\caption{\label{fig:fcst:fwhm_es} Left: same as Fig.~\ref{fig:fcst:ce}, but for the forecast of PSF FWHM.
Right: same as Fig.~\ref{fig:fcst:ce}, but for the logarithm of the radius encircling 80\% of the energy.}
\end{figure}

The absolute differences between cumulative distributions~$F$ of the forecast for several time advances (or image integration time) and the reference normal distribution with zero mean and unit variance are given in Figs.~\ref{fig:fcst:ce}--\ref{fig:fcst:fwhm_es}, where~$S\left[\ln \gamma_1\right]\equiv\frac{\ln \gamma_1 - \mean{\ln \gamma_1}}{\sqrt{\var{\ln \gamma_1}}}$.
This calculation was carried out for the model \arima{4}{0}{1} with a sample size of 10,000 elements for each $N$.
The time intervals ($N$) from 1 to 25 minutes were simulated.
It can be inferred from the error function definition that bell-shaped curves correspond to small bias of the mean.

One can see in these figures, that the maximal absolute difference between the distributions did not exceed $0.03$ for $\ln\left(\gamma_1\right)$, $0.06$ for $\ln\left( \gamma_2\right)$, and $0.12$ for $\ln \left(\gamma_3\right)$.
Moreover, there was explicit light bias of the mean up to~$0.3 \sigma$ for $\ln \left(\gamma_3\right)$, which can be cancelled either empirically or by adding the members of the series expansion in~(\ref{eq:approx_mean_gamma_j}) and (\ref{eq:approx_var_gamma_j}).
When the same numerical experiment was carried out with the 2009-year data, $\ln\left(\gamma_3\right)$ displayed similar behaviour, thus, the bias should be attributed to the calculation of the mean of $\gamma_3$ using the Taylor series.

\subsubsection{Conditional moments}
This section is based on researching the first and second conditional moments $\mean{x_k-x_0|x_0}$ and $\var{x_k-x_0|x_0}$, where $x_0$ is the initial
element of the sequence.
The type of functional connection between the moments and the initial $x_0$ can be found by considering the differential equations that the stochastic series conform to, namely, equation~(\ref{eq:arima:1}).
Moreover, the moments can be estimated using the available observation data.
The discrepancy between the theoretical and empirical laws could imply that the investigated process cannot be described well by the linear differential equations.

Taking into account the linear form of~(\ref{eq:arima:1}), $x_k$ can be expressed as the following linear combination:
\begin{equation}
x_{k} = \sum_{j=1}^{p}x_{j-1} \alpha^{(k)}_{j-1} + \nu_k,
\label{eq:validation:1}
\end{equation}
where $\alpha^{(k)}_{j-1}$ are real coefficients and $\nu_k$ is the linear combination of the generating sequence $a_{i}$.
Applying the conditional mean operation to both parts of~(\ref{eq:validation:1}), given $\mean{\nu}=0$, the following expression is obtained:
\begin{equation}
\mean{x_k-x_0|x_0} = C_1 + C_2 x_0,
\label{eq:validation:2}
\end{equation}
where $C_1$ and $C_2$ are some real coefficients.
Moreover, the connection type does not depend on the specific parameters of the linear model.

It is known that in the linear case the conditional variance of $x_k$ does not depend on initial values, but only on the model parameters~\cite{Box1964}.
Thus, $\var{x_k-x_0|x_0} = C_3$, where $C_3$ is the real constant.

\begin{figure}[t]
\centering
\begin{tabular}{cc}
\includegraphics[height=2.2in]{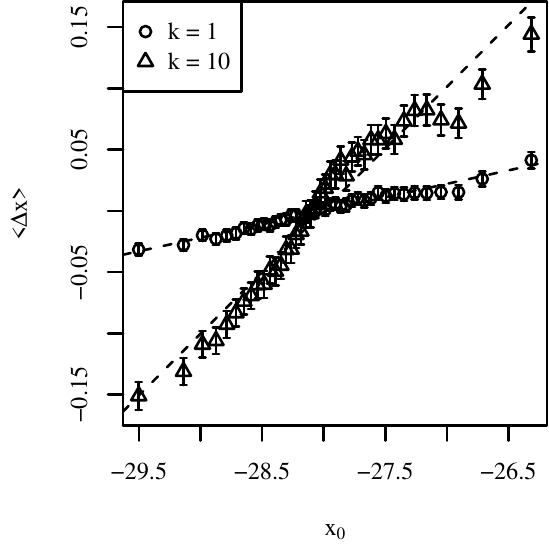} &
\includegraphics[height=2.2in]{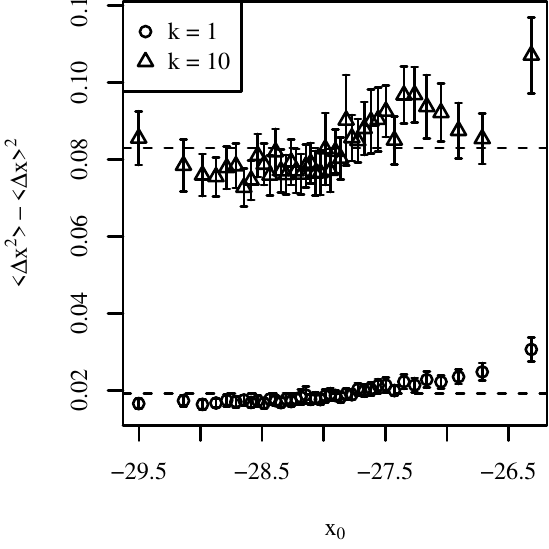}
\end{tabular}
\caption{\label{fig:diff_moment} Sample conditional moments $\mean{x_k-x_0|x_0}$~(left) and $\var{x_k-x_0|x_0}$~(right) for $k=1$ and $k=10$. Note that $x=\ln \left(J\right)$, where $J$ is in units of $\mbox{m}^{1/3}$. }
\end{figure}

The dependence of $\mean{x_k-x_0|x_0}$ and $\var{x_k-x_0|x_0}$ are plotted in Fig.~\ref{fig:diff_moment} for $k=1$ and $k=10$.
It can be seen from these graphs that the calculated conditional moments are well enough described by the linear model in general.
Probably, further model improvements could be made by introducing new atmospheric quantities, but not in regards to the non-linearity.

\section{Conclusions}
\label{sec:outro}

An analysis of the OT data above Mount Shatdzhatmaz, the site of the new MSU telescope, has been carried out in the present paper in order to construct the forecast of the OT characteristics and to schedule online observation tasks in the future.
It has been demonstrated that the OT intensity variation over time on one-minute timescales can be well described by linear auto-regression models, and the parameters of those models were quantified.
The attempt to extract separate nights from the time series was not preferable to the method for simplest model in Section~\ref{sec:demean}.

Overall, a new scheme to forecast the quantities characterising images obtained by an ideal telescope equipped with an ideal detector with long (greater than minutes) exposure times has been proposed.
These quantities are the FWHM, central intensity of the PSF, and radius encircling 80\% of the energy.
The distribution of these parameters has been found to be close to log-normal.
Thus, the calculation of the quantity confidence intervals, which are no less important, is possible.

It has been shown that it is possible to construct the forecast with time advances up to several hours with standard deviations within a few tenths of an arcsecond~(see Fig.~\ref{fig:fcst:variance}).
While forecasting the seeing with a time advance of one hour, the standard deviation was $1.8$ times less than the unconditional standard deviation.
Additionally, it was $1.4$ times less than that for the time advance of two hours.

In summery, a simple model for forecasting image characteristics using the atmospheric OT data has been proposed.
This model in conjunction with other required data will be used to schedule observation tasks online.
The confidence intervals knowledge will make it possible to estimate confidence intervals for the derived quantities such as required exposure times or values of the scheduling cost function, and this should help us to choose more optimal solution-finding algorithms.

\begin{acknowledgements}
The author is grateful to all the people of the MASS group at Sternberg Astronomical Institute.
Moreover, the author would like to give special thanks to B.~Safonov and V.~Kornilov for the valuable discussions on early drafts of this paper.
Lastly, the author highly appreciates the efforts of the anonymous reviewer to make the present paper more understandable to readers.
\end{acknowledgements}

\bibliographystyle{spmpsci}
\bibliography{biblio,biblioi}

\end{document}